\documentclass[twocolumn,showpacs,showkeys]{revtex4}
\usepackage{graphicx}
\begin{document}
\renewcommand\baselinestretch{1.5}

\title{On possible $S$-wave bound states for a $N \bar N$ system
within a constituent quark model}

\author{Chao-Hsi Chang$^{a,b}$ and Hou-Rong Pang$^{b}$}
\affiliation{$^a$CCAST (World Laboratory), P.O.Box 8730, Beijing
100080, China.\footnote{Not correspondence
address.}\\
$^b$Institute of Theoretical Physics, Chinese Academy
of Sciences, P.O.Box 2735, Beijing 100080, China.}

\begin{abstract}
We try to apply a constituent quark model (a variety chiral
constituent quark model) and the resonating group approach for the
multi-quark problems to compute the effective potential between
the $N\bar{N}$ in $S$-wave (the quarks in the nucleons $N$ and
$\bar{N}$, and the two nucleons relatively as well, are in $S$
wave) so as to see the possibility if there may be a tight bound
state of six quarks as indicated by a strong enhancement at
threshold of $p\bar{p}$ in $J/\psi$ and $B$ decays. The effective
potential which we obtain in terms of the model and approach shows
if the experimental enhancement is really caused by a tight
$S$-wave bound state of six quarks, then the quantum number of the
bound state is very likely to be $I=1, J^{PC}=0^{-+}$.
\end{abstract}

\pacs{12.39.-x,13.75.-n,14.20.-c}

\keywords{nucleon-antinucleon effective interaction, multiquark
system, constituent quark model} \maketitle

\section{introduction}
Recently the BES collaboration in the radiative decay $J/ \psi
\rightarrow \gamma p \bar p$ observed a sharp enhancement at
threshold in the $p \bar{p}$ invariant mass spectrum\cite{bai}.
They, further based on a multi-quark bound state conjecture, tried
to fit the enhancement by means of an S-wave Breit-Wigner
resonance function, and obtained the resultant mass peak at
$M=1859^{+3}_{-10}$(stat)$ ^{+5}_{-25}$(sys) MeV and width
$\Gamma$ less than 30 MeV. In the meanwhile, Belle also reported
their observations that in the decays $B^+\rightarrow K^+ p \bar
p$ \cite{k1} and $\bar{B^0} \rightarrow D^0 p \bar p$ \cite{k2},
an enhancement in the $p\bar p$ invariant mass distribution near
the threshold was also seen clearly.

Theoretical investigations of baryon-antibaryon bound states date
back to the proposal of Fermi and Yang\cite{fermi} to make the
pion with a nucleon-antinucleon pair. The extensive and excellent
reviews are given in Ref \cite{richard}. In traditional nuclear
interaction theories within the potential framework which are
based on the single meson exchange mainly, it is shown that $N
\bar N$-system is more attractive than $NN$-system due to the fact
that in the theories there is a strong $\omega$-exchange.
Therefore possible bound states or resonances of a
nucleon-antinucleon system (the so-called nuclear baryonia) have
been proposed for many years.

Recently being encouraged by the observations and with intuition,
several interpretations on the observations near and below the
baryon-antibaryon threshold were suggested: (i) Based on the fact
that the radiative $J/\psi$ decay may be a gluon-rich process, L.
Rosner\cite{rosner} explained the enhancement is due to the
iso-scalar state with $J^{PC}=0^{\pm +}$ being coupled to a pair
of gluons; (ii) B.S. Zou et al with a K-matrix approach \cite{zou}
summed up the one-pion-exchange final state interaction and showed
an important contribution to make the enhancement behavior near
the threshold; (iii) Many phenomenological models were used to
explain this anomalous (sharp) enhancement as a bound state or a
resonance, e.g., A. Datta et al \cite{datta} accounted it for a
new bound state in a simple potential model with a
$\lambda\cdot\lambda$ confining interactions; X.-G. He et
al\cite{xiao} also gave their explanation by using linear $\sigma$
model.

The traditional interaction theories to study $N \bar N $ system,
such as boundary condition model, absorptive potential model,
optical model and coupled-channel models etc, specially put
forward their own method to handle the short-range part of $ N
\bar N $, although in comparatively long range they are quite
similar. For instance, the optical model
\cite{paris1,paris2,paris3,paris4} provides a realistic picture of
the scattering process by introducing an imaginary part of the
potential to reproduce the effect of other excitation channels,
and the spin-isospin dependence of meson-exchange and channel
dependence in the annihilation potential as well are also taken
into account. The coupled-channel model either describes
annihilation in terms of baryon exchange with the same
baryon-meson coupling as in the Yukawa potential \cite{bonn} or
uses semi-phenomenological potential adding partial-wave-analysis
to study $ N \bar N $
interaction\cite{nijmegen1,nijmegen2,nijmegen3}. Although the
agreement with collision experiments is obtained, from the QCD
point of view each of them has shortcoming respectively. For
example, it is hard to imagine that a baryon-exchange picture can
be applied to such a short-range where quarks are `overlapped' and
the color octet configuration must be considered.

In fact, the early studies within the traditional meson exchange
(mesons exchange between the nucleons as whole) framework found
that, if neglecting annihilation channels i.e. taking the real
part of the effective potential only, many bound states might be
formed, while annihilation effects i.e. the imaginary part of the
potential, were included, the binding force decreased and some
bound states were washed out \cite{boundstate}. Moreover, in the
earlier days experimentally the data there was no clear evidence
to imply the existence of strongly bound states. Therefore, we
suspect that the bound state should be tight i.e. the two
components have great overlapping, if a pair of $ N \bar N $
really form a bound state as indicated by BES and Belle
experiments, i.e., the picture with meson-exchanges between the
`whole' nucleons to describe the interaction may not contain all
of the key effects, at least, the behavior of $ N \bar N $
interaction in short-range should be re-considered carefully with
a novel way.

A possible way out for the tight bound state problem is Quantum
chromodynamics (QCD), the fundamental strong interaction physics.
Namely we should start with the quark level, rather than that of
the baryon level. QCD has been already proved to be the right
approach at high energy. At low energy, because of the
non-perturbative nature of QCD, one has to rely on effective
theories underlining QCD and/or QCD-inspired models to get some
insight into the phenomena of the hadronic world. The constituent
quark model is one of them. It has achieved a lot of successes in
describing spectrum for single baryon, the baryon-baryon
interactions and the binding of two baryons such as the nucleus
deuteron, even having found possible strong binding of
$\Omega^-\Omega^-$ systems \cite{zhang,fujiwara,va,wang,zhang2}
etc. Therefore, expanding the constituent quark model into $ N
\bar N $ system is an interesting tentative.

In fact, in order to fitting all of the data from single baryon
spectrum to interactions between two baryons, in literature the
constituent quark model has variety versions. In this paper, to
study of the possibility of $N \bar N$ bound states as indicated
in $J/\psi \rightarrow \gamma p \bar p$, $B^+\rightarrow K^+ p
\bar p$ and $\bar {B^0} \rightarrow D^0 p \bar p$ decays within
the framework of resonating group method, we take a variety chiral
quark model, which may fit the data of single baryon and deuteron
etc. Besides including $\pi$, $\sigma$ and one-gluon exchange, in
the concerned model, we take into account the contributions from
the annihilations of a pair quarks into a meson accordingly and
into one gluon as well. In order to keep the model well-described
for baryon spectrum and $NN$ scattering data, all of the model
parameters have been fixed as possible as those related to $NN$
interaction and baryon spectrum etc.

The paper is organized as follows. In Section II, we derived the
formalism, then gave the Hamiltonian and wave function, after
that, gave a brief summary about resonating group method, which is
used to do dynamical calculation. In Section III, we show the
results and gave discussion.

\section{Hamiltonian,wave function and resonating group method}

\subsection{Formalism}

We think the basis of the constituent quark model for light flavor
systems would be better to set on effective chiral quark model
(with necessary extensions) of QCD. Namely the dynamics should be
described by the effective chiral quark Lagrangian
(pseud-goldstone particles and light quarks as active degrees of
freedom in the theory) \cite{georgi}:
\begin{eqnarray}
L&=&\bar \psi_q (i \not\!\!{D}+\not\!\!{V}+g_A\not\!\!{A}\gamma_5
-m)\psi_q
\nonumber \\
&+& \frac{1}{4}f^2tr\partial^\mu\Sigma^+\partial_\mu\Sigma
-\frac{1}{2}F_{\mu\nu}F^{\mu\nu}+\cdots  \label{(eq1)}
\end{eqnarray}
where
\begin{eqnarray}
V_\mu=\frac{1}{2}(\xi^+\partial_\mu\xi&+&\xi\partial_\mu\xi^+)\;,
\label{(eq2)}\\
A_\mu=\frac{1}{2}(\xi^+\partial_\mu\xi&-&\xi\partial_\mu\xi^+)\;;
\label{(eq3)} \\
\Sigma=e^{2i\Pi/f}\;,\;\;\;\;\;\; \xi&=&e^{i\Pi/f}\;, \;\;\;\;\;\;
(\Sigma=\xi\xi)\;; \label{(eq4)}
\end{eqnarray}
\begin{equation}
\Pi=\frac{1}{2}\Biggl[ {\begin{tabular}{ccc}
$\sqrt{\frac{1}{2}}\pi^0+\sqrt{\frac{1}{6}}\eta$ & $\pi^+$ &$ K^+$\\
$\pi^-$ & $-\sqrt{\frac{1}{2}}\pi^0+\sqrt{\frac{1}{6}}\eta$  & $K^0$\\
$K^-$ & $\bar{K}^0$ & $-\sqrt{\frac{2}{6}}\eta $ \label{(eq5)}\\
\end{tabular}}
\Biggr]\,.
\end{equation}

Since at present we consider the $N\bar{N}$ system, so for the
Lagrangien we restrict ourselves to consider two flavors only. As
mentioned in Introduction, in order to fit nuclear data some
`modification' on the quark chiral model is needed, such as to add
a scalar $\sigma$ `meson' \cite{pdg} and a `survived' effective
gluon into the model etc. Thus with the modification the relevant
quark-gluon-meson Lagrangian can be re-written as
\begin{equation}
L_{ch}=\bar \psi_q (i\not\!\!D-m_q) \psi_q +g_{ch}\bar \psi_q
(\sigma+i \gamma_5 \overrightarrow \tau \cdot \overrightarrow
\pi)\psi_q\,,  \label{eq6}
\end{equation}
where $\psi_q$ denotes quark field, and $m_q\; (q=u, d)$ are the
constituent quark masses. Current masses turns to constituent
masses accordingly is a consequence of the chiral symmetry
breaking. $g_{ch}$ is the vertex coupling (quarks to mesons)
constant and $D^\mu=\partial^\mu+ig_s\frac{{\lambda^a}}
{2}A^{a\mu}$ ($a$ is an index of color space and $\lambda^a$ is
Gell-Mann matrix for $SU_c(3)$ color) $A^{a\mu}$ is gluon field.
To study the baryon structure and baryon-baryon interaction, the
survived gluon field $A^{a\mu}$ is introduced so as to take care
of the necessary non-perturbative contributions in the model. In
addition, to provide the non-perturbative QCD effects at long
distance, an effective confinement in potential is needed. We will
discuss them later.

Here when computing the effective potential between $N$ and
$\bar{N}$, we only take into account the tree Feynman diagrams:
the relevant exchange and annihilation diagrams (see Fig.1-3).
\begin{figure}
\includegraphics[width=0.3\textwidth,angle=0]{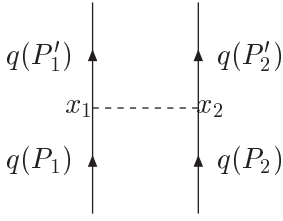}
\vspace{-1.0cm}\caption{The exchange diagrams between two quarks.}
\label{FIG1}
\end{figure}
\vspace{-0.5cm}
\begin{figure}
\includegraphics[width=0.3\textwidth,angle=0]{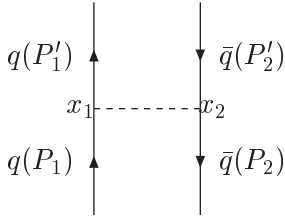}
\vspace{-1.0cm} \caption{The exchange diagrams between quark and
anti-quark.} \label{FIG2}
\end{figure}
\vspace{0.5cm}

The effective potential derived from the `exchange' between quarks
has been considered by many authors such as those of
Refs.\cite{zhang,fujiwara,va,wang,zhang2}, now we should consider
those derived from the exchange and annihilation between quark and
antiquark.

(i). The effective potential derived from `exchange':

According Feynman rule, the T-matrix of $\sigma$-exchange between
quark and anti-quark can be written as
\begin{equation}
\bar u(p'_1,s'_1) u(p_1,s_1)
\frac{-F^2(\overrightarrow{q}^2)}{q^2}g^2_{ch}
  \bar v(p_2,s_2)v(p'_2,s'_2) \,.\label{(eq7)}
\end{equation}
Here, we have inserted a form factor $F(\overrightarrow{q}^2)$ to
cut off the contributions from the short-distance region where the
chiral symmetry has not broken. Generally, it takes the form
$F(\overrightarrow{q}^2)=(\frac{\Lambda
^2}{\Lambda^2+\overrightarrow{q}^2})^{1/2}$. $p_1, s_1, p'_1,
s'_1$ are the four-vector momenta and spin $z$-projections of
initial quark and final quark, respectively. $p_2, s_2, p'_2,
s'_2$ are the four-vector momenta and spin $z$-projections of
initial antiquark and final antiquark, respectively. $u$ and $v$
are assumed to be free Dirac spinors for quarks and antiquarks,
respectively, but soon in the paper we will replace them by two
component Pauli spinors for convenience because of the
non-relativistic nature for the problem.

When we take the non-relativistic limit to turn the Dirac spinors
to Pauli ones, there is an additional minus sign from the
interchange of the effective interaction $\bar{u}u\bar{u}u$ to the
$\bar{u}u\bar{v}v$ so it cancels the extra fermion minus sign.
Therefore the $\sigma$-exchange potential is universally
attractive, no matter between a pair of fermions or between a pair
of antifermions (or a pair of each). The $\sigma$-exchange
potential between quarks or quark-antiquark in the nonrelativistic
problem in coordinate space can been written as
\begin{equation}
V^{\sigma}_{qq}(r) = V^{\sigma_{exch}}_{q\bar q}(r) =
-\frac{g^2_{ch}}{4\pi}\frac{\Lambda^2}{\Lambda^2 - m_{\sigma}^2}
[\frac{e^{-m_\sigma r}}{r}-\frac{e^{-\Lambda r}}{r}]\,.  \label{(eq8)}
\end{equation}

As for the $\pi$-exchange, since it has a different sign when
taking non-relativistic limit from Dirac spinors to Pauli ones for
the effective interactions $\bar{u}\gamma^5 u \bar{u} \gamma^5 u$
and $\bar{u} \gamma^5 u \bar{v}\gamma^5 v$, so there is an
additional minus sign from fermion-fermion into
fermion-antifermion in taking the limit.
\begin{eqnarray}
V^{\pi_{exch}}_{q\bar q}(r) &=& -\frac{1}{12}\frac{g^2_{ch}}{4\pi}
\frac{m^2_\pi}{m_{q_i}m_{q_j}}[\frac{e^{-m_\pi
r}}{r}-\frac{\Lambda^2}{m^2_\pi} \frac{e^{-\Lambda r}}{r}] \nonumber \\
&&\frac{\Lambda^2}{\Lambda^2-m_\pi^2} (\overrightarrow{\sigma_i}
\cdot \overrightarrow{\sigma_j})(
\overrightarrow{\tau_i} \cdot \overrightarrow{\tau_j})\,; \nonumber\\
V^{\pi}_{qq}&=&\frac{1}{12}\frac{g^2_{ch}}{4\pi}
\frac{m^2_\pi}{m_{q_i}m_{q_j}}[\frac{e^{-m_\pi
r}}{r}-\frac{\Lambda^2}{m^2_\pi} \frac{e^{-\Lambda r}}{r}] \nonumber \\
&&\frac{\Lambda^2}{\Lambda^2-m_\pi^2} (\overline{\sigma_i} \cdot
\overrightarrow{\sigma_j})(\overrightarrow{\tau_i} \cdot
\overrightarrow{\tau_j})\,. \label{(eq9)}
\end{eqnarray}
We note that the additional minus may be understood by the
G-parity rule easily. Here $\overrightarrow{\sigma}_{i(j)}$ is a
Pauli matrix for spin and $\overrightarrow{\tau}_{i(j)}$ is a
Pauli matrix for isospin.

Since it is not allowed any quark-antiquark exchange between $N$
and $\bar{N}$, so the interaction from one-gluon-exchange for $N$
and $\bar{N}$ is very different that for the $N N$ system that
one-gluon-exchange between $N$ and $\bar{N}$ does not contribute
to the effective potential at all. Here we only write down the
potential due to one-gluon exchange between two quarks (antiquark)
in one baryon (antibaryon)\cite{georgi},
\begin{eqnarray}
&&V^g_{qq(\bar{q}\bar{q})}(r)\propto \nonumber \\
&&\frac{(\vec{\lambda}_i \cdot \vec{\lambda}_j)}{4} \left[
\frac{1}{r}-\frac{\pi\delta (\vec{r})}{2} \left(
\frac{1}{m^2_i}+\frac{1}{m^2_j}+\frac{4(\vec{\sigma}_i \cdot
\vec{\sigma}_j)}{3m_im_j} \right) \right]\,.\label{(eq10)}
\end{eqnarray}

(ii). The effective potential derived from annihilation:

Let us take $\pi$ as an example, and for it the T-matrix of
$\pi$-annihilation can be written as
\begin{equation}
-\bar u(p'_1,s'_1)\gamma^5 v(p'_2,s'_2) \frac{-1}{q^2}g^2_{ch}
\bar v(p_2,s_2)\gamma^5u(p_,s_1)\,.\label{(eq11)}
\end{equation}
\begin{figure}
\includegraphics[width=0.25\textwidth,angle=0]{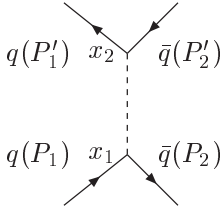}
\vspace{-1.2cm} \caption{The annihilation diagrams for quark and
anti-quark.} \label{FIG3}
\end{figure}\vspace{-0.4cm}

Using Fierz identities for Dirac matrices, a four-Fermion operator
can be expressed definitely as a linear superposition of others
with a changed sequence of spinors as follows.
\[(\bar a O_i b)(\bar c O^i d)= \sum_{k=1,16}C_{ik} (\bar a O_k
d)(\bar c O^k b)\,. \]

Since we are taking into account the contributions of the lowest
order to the `full' $S$-wave at this step, i.e., we consider the
`full' $S$ wave cases (all of the quarks and antiquarks are in
$S$-wave), so under static approximation the contributions from
$\sigma$-meson ($J^{PC}=0^{++}$) annihilation ($P$-wave
annihilation) to the potential in the present case (for $S$-wave
states) are tiny. Thus we omit them here. The contributions from
one-gluon annihilation to the potential between quark and
anti-quark in momentum representation can be written as
\begin{eqnarray}
&V^g_{anni}=\displaystyle \frac{4\pi \alpha'_s}{s}\frac{1}{4}
\Bigg(16/9-1/3 (\overrightarrow{ \lambda^c_i}\cdot
\overrightarrow{
\lambda^c_j})\Bigg) \nonumber \\
&\Bigg(1/2+1/2 (\overrightarrow{ \tau_i}\cdot \overrightarrow{
\tau_j})\Bigg) \Bigg(-3/2+1/2 (\overrightarrow{ \sigma_i}\cdot
\overrightarrow{ \sigma_j})\Bigg)\,.\label{(eq12)}
\end{eqnarray}
Here $s=(p_1+p_2)^2$ (in 4-dimension) in the propagator.

In an annihilation process, $s$ is time-like ($s > 0$), under
static approximation so we have $s\simeq(m_1+m_2)^2$, and when
transferring the potential to space-time representation, we have
\begin{eqnarray} &&V^g_{anni}= -\frac{4\pi
\alpha'_s \delta(r)}{(m_i+m_j)^2}\frac{1}{4} \Bigg(16/9-1/3
(\overrightarrow{
\lambda^c_i}\cdot \overrightarrow{
\lambda^c_j})\Bigg) \nonumber \\
&&\Bigg(1/2+1/2 (\overrightarrow{
\tau_i}\cdot
\overrightarrow{
\tau_j})\Bigg) \Bigg(-3/2+1/2 (\overrightarrow{
\sigma_i}\cdot \overrightarrow{
\sigma_j})\Bigg)~~\,. \label{(eq13)}
\end{eqnarray}

For $\pi$ annihilation, we have
\begin{eqnarray}
&V^{\pi}_{anni}&=\frac{ g^2_{ch} \delta(r)}{m_{\pi}^2-(m_i+m_j)^2}
\Bigg(-1/2-1/2(\overrightarrow{\sigma_i}\cdot\overrightarrow{
\sigma_j})\Bigg)\nonumber \\
&&\Bigg(3/2-1/2(\overrightarrow{\tau_i}\cdot\overrightarrow{
\tau_j})\Bigg)
\Bigg(1/3+1/2(\overrightarrow{\lambda^c_i}\cdot\overrightarrow{
\lambda^c_j})\Bigg)~~~~~\,. \label{(14)}
\end{eqnarray}

\subsection{Hamiltonian and the wave functions of the model}

We follow the chiral quark model for multi-quark systems
\cite{zhang2}, which essentially is an effective theory on
exchanges of Goldstone mesons, scalar meson $\sigma$ and gluons as
well between quarks, but we extend it to antiquarks involved, so
as to study the nucleon-antinucleon system with definite isospin
(I) and spin (S).

As the first step and to the conjecture of BES collaboration
\cite{bai}, here only S-wave states of the nucleon pair are
considered, i.e., the total orbital angular momentum $L=0$, and we
have $J=S$ (the total angular momentum comes from quark spin
only). The most parameters are fixed by fitting baryon spectrum of
the model\cite{zhang2}: The coupling constant $g^2_{ch}/4\pi$ is
fixed by $g^2_{NN\pi}/4\pi$, i.e., $g^2_{ch}/4\pi$ is related to
$f^2_{qq\pi}$ directly by $\frac{m^2_\pi}{4m_Nm_N}g^2_{ch}/4\pi
=f^2_{qq\pi}$, ($m_\pi,m_N$ are the masses of pion and nucleon,
respectively), the one-gluon-exchange coupling constant $\alpha_s$
is determined by mass splitting of $\Delta$ and $N$, the
confinement strength $a_c$ is fixed by the stability conditions of
$N$, $V_0$ is fixed by the masses of $N$. In summary, the
parameters in the model are listed in Table I. The units for $m_q,
m_\sigma, m_\pi, V_0, b$ and $a_c$ are MeV, fm and $MeV\cdot
fm^{-2}$ respectively.

\vspace{4mm} Table I. The parameters of the model.
\begin{center}
\begin{tabular}{ccccccccc} \hline\hline
$m_q$&$b$&$\alpha_s$&$a_c$&$\frac{g^2_{ch}}{4\pi}$
&$m_{\sigma}$&$m_{\pi}$&$\Lambda$&$V_0$\\
\hline 313.~&~0.60~&~0.95~&~7.3~&~0.59~
&~570.~&~138.~&~829.~&~72.5~
\\ \hline \hline
\end{tabular}
\end{center}
To extend this model from $NN$ systems to $N\bar{N}$
(nucleon-antinucleon) systems based on the effective Lagrangian
Eq.(\ref{eq6}), now the Hamiltonians of the model may be written
as
\begin{eqnarray} H_{p\bar p} &=& \sum_{i=1}^6
(m_i+\frac{p_i^2}{2m_i})-T_{CM} \nonumber \\
&&+\sum_{i=1,2,3\hspace{5pt}
j=4,5,6} V(r_{ij}) \,,   \label{(eq15)} \\
&&V(r_{ij})=V_{conf}(r_{ij}) + V^{exch}_{q\bar q}(r_{ij})
+V^{anni}_{q\bar q}(r_{ij})]\,, \nonumber  \\
&&V^{exch}_{q\bar q} = V^{\sigma_{exch}}_{q\bar q}(r_{ij})
+V^{\pi_{exch}}_{q\bar q}(r_{ij})\,;\nonumber \\
&&V^{anni}_{q\bar q}(r_{ij})= V^{\pi_{anni}}_{q\bar q}(r_{ij})
+V^{g_{anni}}_{q\bar q}(r_{ij})\,;\nonumber \\
&&V_{conf}(r_{ij})= -a_c\lambda_i^c \cdot\lambda_j^c r_{ij}^2 +
V_0\,,\nonumber
\end{eqnarray}
where $a_c$ is the confinement strength. $T_{CM}$ is the kinetic
energy in the system of center mass. $V^{\pi_{exch}}_{q \bar q}$
and $V^{\sigma_{exch}}_{q \bar q}$ are the effective potential
from $\pi$ and $\sigma$ exchanges between the couple of quark and
antiquark, respectively. $V^{\pi_{anni}}_{q \bar q}$ and
$V^{g_{anni}}_{q \bar q}$ are those from the $\pi$ and one-gluon
annihilation, respectively.

Each of the wave functions for nucleon and antinucleon can be
written as products of three parts respectively:
\begin{eqnarray}
\Phi_p=\Phi^O_p \Phi^{sf}_p\Phi^c_p \,;\;\;\;\; \Phi_{\bar
p}=\Phi^O_{\bar p} \Phi^{sf}_{\bar p}\Phi^c_{\bar p}\,.\nonumber
\end{eqnarray}
For the orbital and color parts, the proton and antiproton have
the same `internal motion' wave function (here the motion for each
center of mass has been removed):
\begin{eqnarray}
&\Phi^O_p= (\frac{2}{3\pi b^2})^{3/4}(\frac{2}{4\pi b^2})^{3/4}
e^{-(\lambda_1^2/(3b^2)+\rho_1^2/(4b^2))},\nonumber \\
&\Phi^O_{\bar p}=(\frac{2}{3\pi b^2})^{3/4}(\frac{2}{4\pi
b^2})^{3/4} e^{-(\lambda_2^2/(3b^2)+\rho_2^2/(4b^2))},\nonumber
\end{eqnarray}
here $\lambda,\rho$ are the Jacobi coordinates of the components
in each of the two clusters (nucleon, anti-nuclen), respectively.
For the color factors:
\begin{eqnarray}
&\Phi^c_p=\frac{1}{\sqrt{6}}(ryb-rby+ybr-yrb+bry-byr),\nonumber\\
&\Phi^c_{\bar p}=\frac{1}{\sqrt{6}}(\bar r \bar y \bar b- \bar r
\bar b \bar y+\bar y \bar b \bar r -\bar y \bar r \bar b+\bar b
\bar r \bar y- \bar b \bar y \bar r)\,. \nonumber
\end{eqnarray}

For the flavor factors for a $N\bar{N}$ system, there are four
possibilities with definite quantum numbers I and J. They are
precisely (all symbols here have their usual meanings):
\begin{eqnarray}
&\frac{1}{2}(p\uparrow\bar{p}\downarrow +p\downarrow
\bar{p}\uparrow- n\uparrow \bar{n}\downarrow -n\downarrow
\bar{n}\uparrow) \nonumber\\
&for\;\;\; I,J^{PC}=1,1^{--}\,; \nonumber\\
&\frac{1}{2}(p\uparrow \bar p \downarrow -p\downarrow \bar p
\uparrow- n\uparrow \bar n \downarrow +n \downarrow \bar n
\uparrow ), \nonumber\\
&for \;\;\; I,J^{PC}=1,0^{-+}\,; \nonumber\\
&\frac{1}{2}(p\uparrow \bar p \downarrow +p\downarrow \bar p
\uparrow+ n\uparrow \bar n \downarrow
+n \downarrow \bar n \uparrow ), \nonumber\\
&for\;\;\; I,J^{PC}=0,1^{--}\,; \nonumber\\
&\frac{1}{2}(p\uparrow \bar p \downarrow -p\downarrow \bar p
\uparrow+ n\uparrow \bar n \downarrow -n \downarrow \bar n
\uparrow ),  \nonumber\\
&for\;\;\; I,J^{PC}=0,0^{-+}\,. \nonumber
\end{eqnarray}

If we relate the states to the observation at BES with the decay
$J/\psi\to p\bar{p}+\gamma$\cite{zhu}, we are sure that the
$C$-parity of the $p \bar{p}$ pair must be positive i.e. $C = +$.
Furthermore if we restrict ourselves to take S-wave into account
only, then the states with minimal total angular momentum can be
$J^{PC}=0^{-+} (L=0, S=0)$ only. Whereas, here we also consider
the states with $J^{PC}=1^{--} (L=0, S=1)$ for comparison.

\subsection{A outline of the resonating group method}

For the the resonating group method \cite{buchmann}, first of all,
to write down the two-cluster wave function with the conventional
ansatz (to factorize out the relative motion of mass centers of
the two `clusters') as follows,
\begin{eqnarray}
 |\Psi_{p\bar p} \rangle = [\Phi_p \Phi_{\bar p}]^{[c]IS}
\chi(\overrightarrow{R})\,. \label{(16)}
\end{eqnarray}
where [c]=[222] gives the total color symmetry.
$\chi(\overrightarrow{R})$ is relative motion wave function of the
two clusters. $\Phi_p$ and $\Phi_{\bar p}$ are the wave functions
of the nucleon and antinucleon clusters in isospin and spin space
only.

To the specific problem, Gaussian functions with various reference
centers $\vec{S}_i$ (i=1...n) are introduced, which ($\vec{S}_i$)
play the `generating coordinates' in the formalism,
$$ \chi_i(\vec{R},\vec{S}_i)=(\frac{3}{2 \pi b^2})^{3/4}exp\{ -\frac{3}{4 b^2}(\vec{R}-
\vec {S_i})^2 \} $$ and the relative motion wave function of the
two clusters of the quarks and antiquarks is expanded into partial
waves
\begin{eqnarray}
\chi(\vec{R})&=& \sum_{L}\chi^L(R)Y^{LM}(\hat{\vec{R}}) \nonumber \\
&=&\sum_{L}\sum^N_{i=1}c_i^L\chi_i^L(R,S_i)Y^{LM}(\hat{\vec{R}})\,,
\label{(eq17)}
\end{eqnarray}
with
\begin{eqnarray}
&&\chi_i^L(R,S_i)\equiv \int
d\Omega_{S_i}\chi_i(\vec{R},\vec{S}_i)Y^{LM}(\hat{\vec{S_i}}) \nonumber \\
&&=(\frac{3}{2 \pi b^2})^{3/4} \int exp\{ -\frac{3}{4
b^2}(\vec{R}-\vec{S_i})^2 \}
Y^{LM}(\hat{\vec{S_i}})d\Omega_{S_i}  \nonumber \\
&&= 4\pi (\frac{3}{2 \pi b^2})^{3/4}
exp\{-\frac{3}{4b^2}(R^2+S_i^2) \}  i_L(\frac{3}{2b^2}RS_i)\,,
\nonumber
\end{eqnarray}
where $i_L$ is the L-th modified spherical Bessel function. For
L=0, one has $i_0(x)=sinh(x)/x$ (for a bound state).

According to the ansatz of the RGM and having the center of mass
motion
$$\Phi_{cm}(\overrightarrow{R}_{cm})=(\frac{6}{\pi b^2})
^{\frac{3}{4}}exp(-\frac{3\overrightarrow{R}_{cm}^2}{b^2})$$
included, finally the wave function of six quarks within the
two-clusters accordingly can be written as
\begin{eqnarray}
&&\Psi_{6q} = {\cal A} \sum_k \sum_{i=1}^{n} C_{k,i}
  \int d\Omega_{S_i}\prod_{\alpha=1}^{3}
  \prod_{\beta=4}^{6}\cdot \nonumber \\
&&\psi_\alpha(\vec{S_i}) \psi_\beta(-\vec{S_i})
    [\Phi_p^{s_1f_1}\Phi_{\bar p}^{s_2f_2}]^{I,J=S}
   [\Phi^c_p\Phi^c_{\bar{p}}]^{[\sigma]}\,, \label{(eq18)}
\end{eqnarray}
here $\psi_\alpha(\vec{S_i})$ and $\psi_\beta(-\vec{S_i})$ are the
single-particle orbital wave functions with different reference
centers \begin{eqnarray} &\psi_\alpha(\vec{S_i}) =(\frac{1}{\pi
b^2})^{3/4}e^{-(\vec{r}_\alpha-\vec{S_i}/2)^2/(2b^2)}\,, \nonumber\\
&\psi_\beta(-\vec{S_i}) =(\frac{1}{\pi b^2})^{3/4}
e^{-(\vec{r}_\beta+\vec{S_i}/2)^2/(2b^2)}\,. \nonumber
\end{eqnarray}
With the variational principle, one may obtain the RGM equation
\begin{equation}
\int H(\vec R, \vec{R'}) \chi (\vec{R'}) d\vec{R'} = E \int N(\vec
R,\vec{R'}) \chi (\vec{R'}) d\vec{R'} \label{(eq19)}
\end{equation}
via the variation with respect to the relative motion wave
function $\chi(R)$. With a re-formulation, the RGM equation
becomes an algebraic eigenvalue equation \cite{buchmann}
\begin{equation}
\sum_{j,k'} C_{j,k'} H_{i,j}^{k,k'} = E \sum_{j} C_{j,k}
N_{i,j}^{k}\,. \label{(eq20)}
\end{equation}
We should note here that for the nucleon-antinucleon system only
the direct terms contribute.

\section{Numerical results and discussions}

There are four possible states for an $S$-wave nucleon-antinucleon
system with different isospin $I$ and total angular momentum $J$
respectively. The spin, isospin and spin-isospin matrix elements
of the interaction for the possible states are listed in Table II.

\vspace{2pt}
\begin{center}
Table II. The spin, spin-isospin and isospin coefficients for
different states
\vspace{2mm}
\begin{tabular}{ccccc} \hline\hline
~~~$I~~J^{PC}$~~~&~$ 1~~ 1^{--}$~~&~$1~~
0^{-+}$~~&~$
0~~ 1^{--}$~&~$ 0~~ 0^{-+}$ \\
\hline $\langle \sigma_i\cdot\bar\sigma_j \rangle$&
-1/9&1/3&-1/9&1/3 \\
$\langle \tau_i\cdot\bar\tau_j
\rangle$& -1/9&-1/9&1/3&1/3 \\
$\langle\sigma_i\cdot\bar\sigma_j \tau_i\cdot\bar \tau_j
\rangle$& 25/81&-75/81&-75/81&25/9 \\ \hline\hline
\end{tabular}
\end{center}
\vspace{1pt}

For a qualitative analysis, firstly we neglect the nonlocal terms
of the potential, so effective nucleon-antinucleon potential can
been obtained by the Born-Oppenheimer approximation. The
dependence of the `potential' (minus the `rest masses' of the
proton and the antiproton) on the separate distance $S_i$  is
defined as the expectation value of the Hamiltonian at a separate
distance $S_i$ as follows,
\begin{eqnarray}
&V_{p\bar p}(S_i)=\displaystyle \frac{\langle \Psi_{p\bar p}(S_i)
| H| \Psi_{p\bar p}(S_i) \rangle }{\langle \Psi_{p\bar p}(S_i) |
\Psi_{p\bar p}(S_i) \rangle }\nonumber \\
&-\langle \Psi_p|H|\Psi_p \rangle -\langle \Psi_{\bar
p}|H|\Psi_{\bar p} \rangle\,. \label{(eq21)}
\end{eqnarray}
To see it precisely, we have calculated the nucleon-antinucleon
effective potential with various gluon annihilation coupling
constant. For simplifying, only the results in the cases with
$I,J=1,0$ are drawn in FIG.4. The contributions from $\pi$ and
$\sigma$-exchange, $\pi$-annihilation and one-gluon annihilation
with different isospin and spin are shown in FIG.5.

In the calculations, all the model parameters are determined by
fitting nucleon-nucleon interaction and deuteron properties (for
the annihilation contribution, the parameters are related to those
for exchanges through the Lagrangian Eq.(\ref{eq6})). From QCD, we
know, one-gluon running coupling constant would become small with
the increasing of momentum transfer. Here we let gluon
annihilation coupling constant $\alpha'_s$ equal to one-gluon
exchange coupling constant as that $\alpha_s$ in Table I. firstly,
then reduce $\alpha'_s$ appropriately to see the dependence on
gluon annihilation coupling constants for the potential of the
nucleon-antinucleon systems. For comparison, the case to ignore
the contributions from annihilations are also computed.

\begin{center}
\begin{figure}
\includegraphics[width=0.35\textwidth]{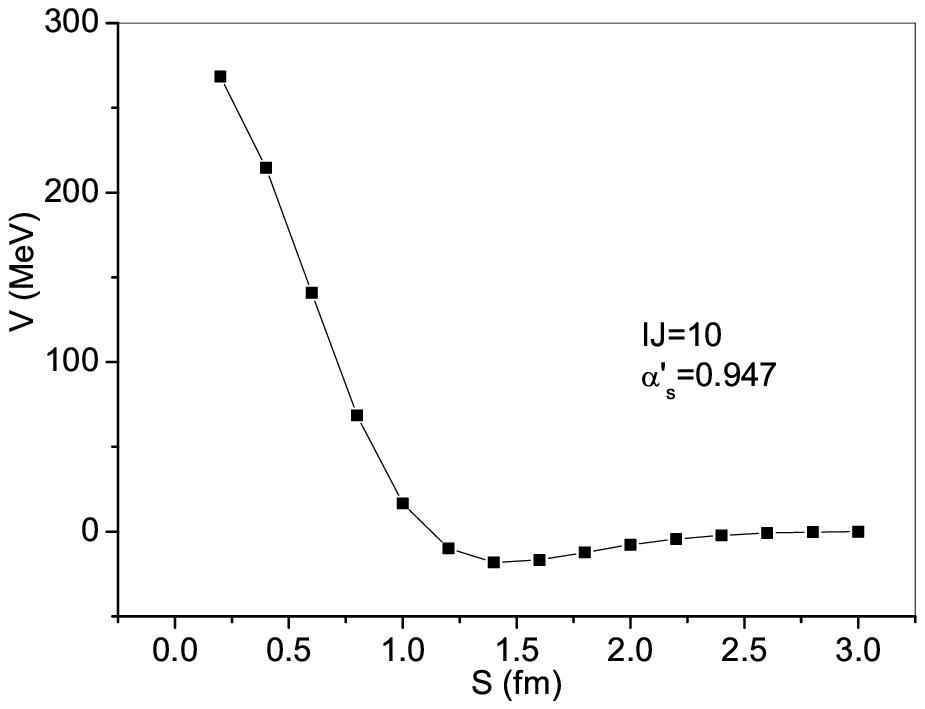}
\hspace{-2.0cm}\includegraphics[width=0.35\textwidth,angle=0]{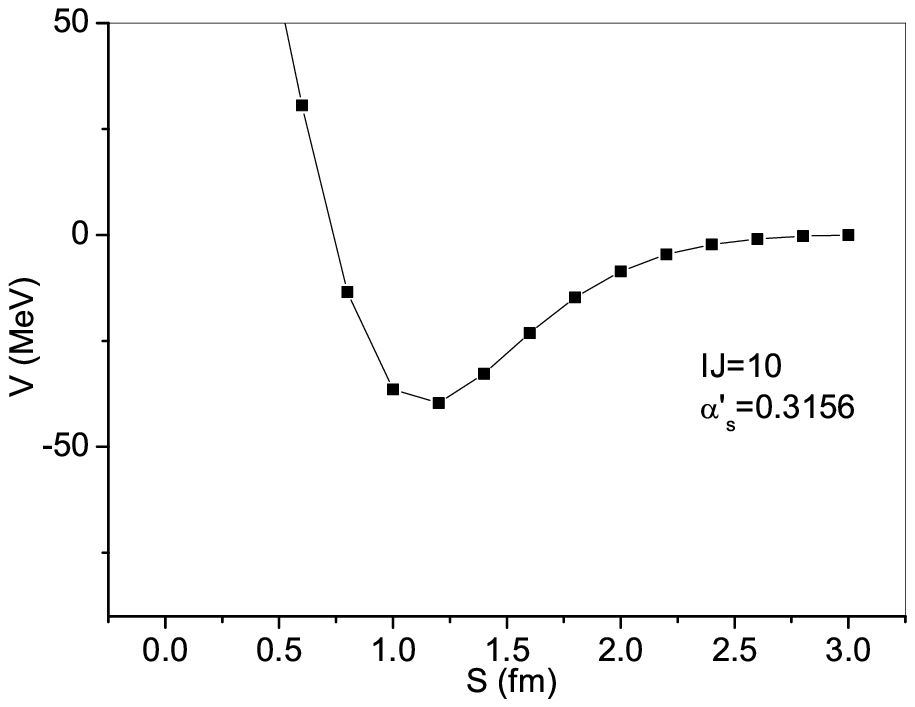}
\vspace{-0.6cm} \caption{The effective potential for an $S$-wave
nucleon-antinucleon system with different gluon annihilation
coupling constant with state $I,J=1,0$.} \label{FIG4}
\end{figure}
\end{center}
\vspace{-1.0cm}

FIG.4 contains 2 sub-figures. It shows the total
nucleon-antinucleon effective potential with
$\alpha'_s=\alpha_s=0.95$ and
$\alpha'_s=\frac{1}{3}\alpha_s=0.316$ for $I,J=1,0$ state. FIG.5
contains 3 sub-figures. It shows the contributions from $\pi$
exchange, $\pi$ annihilation, $\sigma$ exchange and one-gluon
annihilation for the states $I, J^{PC}= 1, 0^{-+};$ $ 0, 1^{--};
0, 0^{-+}$ respectively. For the other states, we only present
their minimum values of the effective 'potential' in Table III.

\vspace{2mm} Table III. List of the minimum of the effective
potential for the possible states (unit for $V_{min}$ in MeV).

\begin{tabular}{ccccc} \hline\hline
IJ  &11&10&01&00   \\ \hline
~$\alpha'_s$=0.947~~~&~~-20.1~~&~~-18.1~~&~~-14.1~~&~~-14.9 \\
\hline ~$\alpha'_s$=0.4735~~~&~~-31.7~~&~~-32.3~~&~~-32.1~~&~~-16.2 \\
\hline ~$\alpha'_s$=0.3156~~~&~~-37.4~~&~~-39.7~~&~~-43.6~~&~~-16.7 \\
\hline without annihilation&~~-89.3~~&-150.6~~&~~-150.6~~&~~-21.4 \\
\hline\hline
\end{tabular}

\vspace{2mm}

If the attraction of the effective potential is deep enough for
forming bound states, we further do dynamical calculations in the
framework of resonating group method exactly, and finally obtain
corresponding relative motion wave functions and the mean square
roots of the radius as well for the states.

From FIG.5 we may see that $\sigma$ exchange gives a universally
attractive interaction for a nucleon and antinucleon system as it
does for a nucleon-nucleon system. As stated before, the
contributions from $\sigma$ annihilation to the $S$-wave nucleon
and antinucleon system is small, so we omit it. $\pi$ annihilation
gives repulsion interaction for all concerned cases with different
strength, the contribution to $I, J^{PC}=1, 0^{-+}$ is the
largest, while the contribution to $I, J^{PC}=0, 0^{-+}$ is the
smallest. One-gluon annihilation also provides rather repulsion
for these four considered states, especially, for the states $I,
J^{PC}=1, 0^{-+}; 0, 1^{--}$. (It seems that the contribution from
the gluon annihilation should not be omitted when studying
nucleon-antinucleon interaction, unless coupling constant of gluon
annihilation is much smaller than that of gluon exchange.) $\pi$
exchange is the main factor to cause the differences for these
four states. It provides repulsion for the state $I, J^{PC}=0,
0^{-+}$ (and a minor attraction in the long range) but attraction
for states $I, J^{PC}=0, 1^{--}; 0, 1^{--}$, that is opposite to
the case for nucleon-nucleon systems.

\begin{center}
\begin{figure}
\includegraphics[width=0.4\textwidth,angle=0]{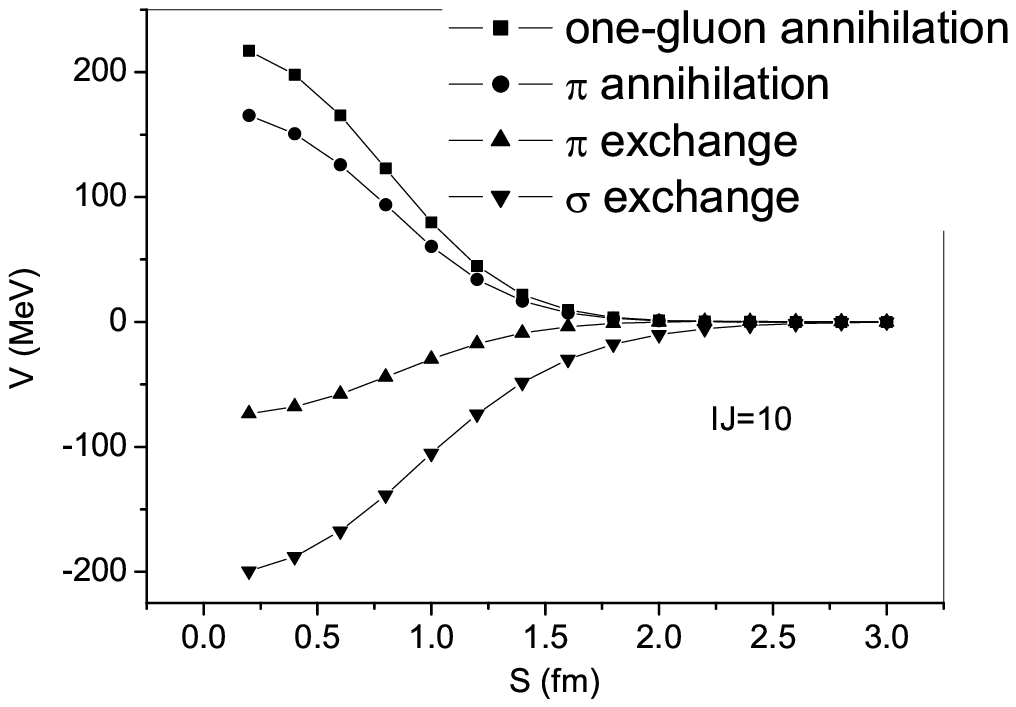}
\includegraphics[width=0.37\textwidth,angle=0]{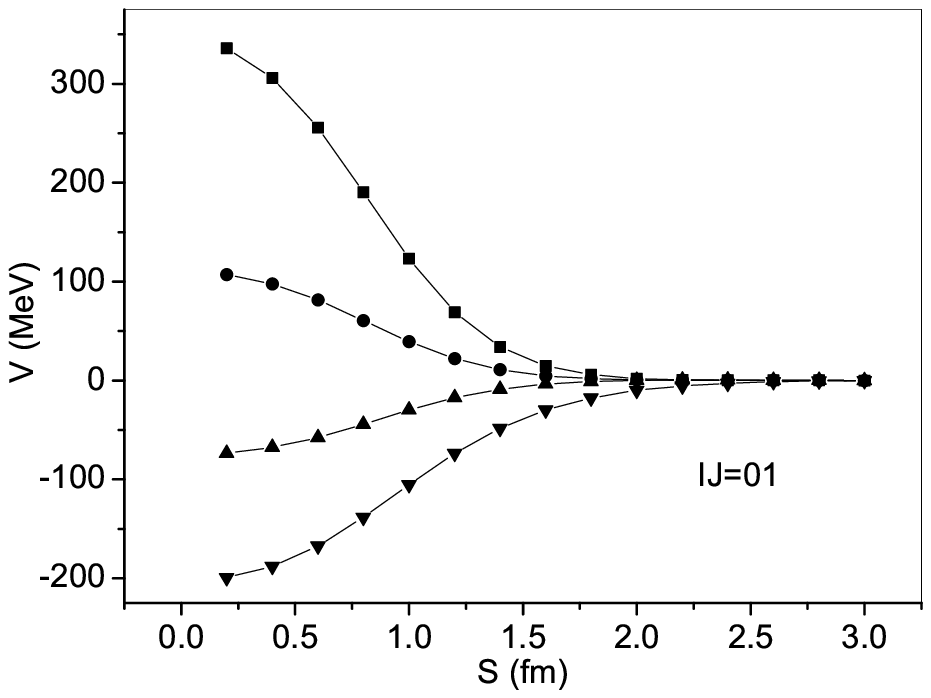}
\hspace{-2.0cm}\includegraphics[width=0.38\textwidth,angle=0]{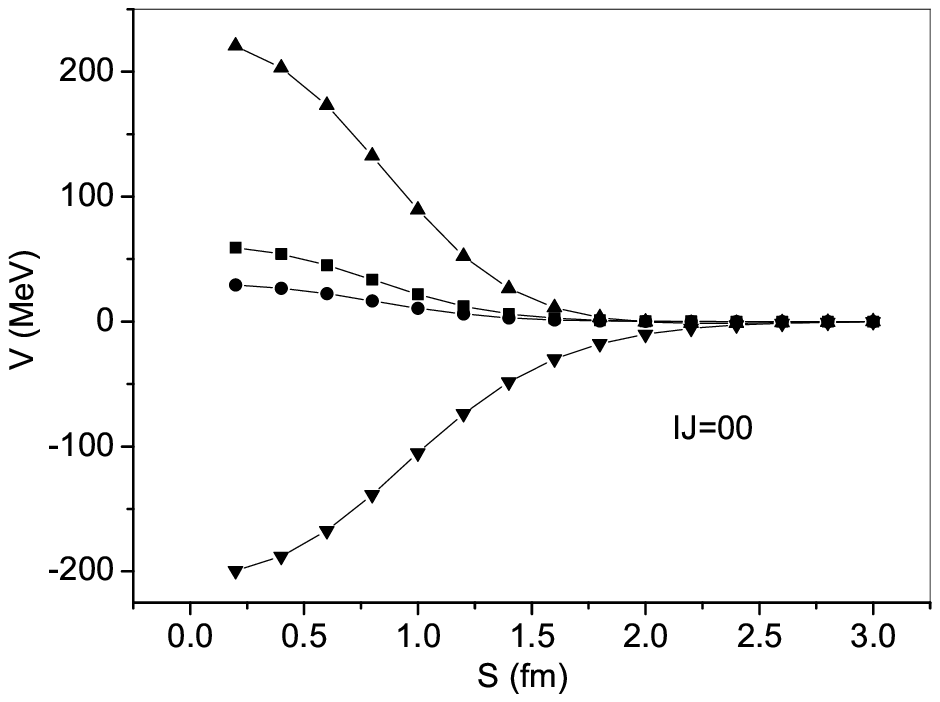}
\vspace{-0.6cm} \caption{The contributions to the effective
potential from $\pi$ and $\sigma$-exchange, $\pi$-annihilation and
one-gluon annihilation for the states $I,J=1,0;\,0,1;\,0,0$.
Square points denote the contribution from gluon annihilation,
circle ones for pion annihilation, up-triangle ones for pion
exchange and down-triangle ones for $\sigma$ exchange.}
\label{FIG5}
\end{figure}
\end{center}
\vspace{-1.0cm}

\begin{center}
\begin{figure}
\includegraphics[width=0.4\textwidth,angle=0]{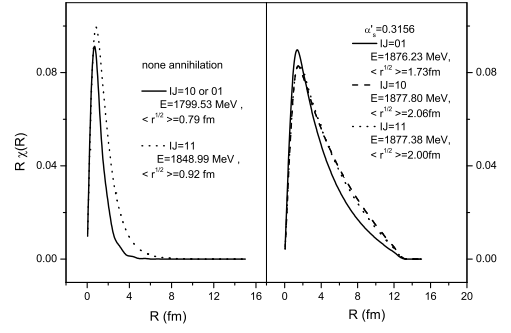}
\caption{Relative motion wave functions for the states
$I,J^{PC}=0,1^{--};\, 1,0^{-+}$ and $1,1^{--}$ with $\alpha'_s
=0.3156$ (the left figure), and without annihilation contributions
at all (the right figure).} \label{FIG6}
\end{figure}
\end{center}
\vspace{-1.2cm}

From Table III, one may see that the four possible states with
different isospin and spin quantum numbers, all have intermediate
attraction with different strength, and the attraction increases
with decreasing the coupling constant $\alpha'_s$ of the gluon
annihilation. The observation by BES collaboration presents a
strong enhancement in the decay $J/\psi \rightarrow \gamma p \bar
p $, that imply the quantum numbers of the $p \bar p$ state, if it
is in S-wave, are $I\,(J^{PC})=0$ or $1\,\; (0^{-+})$ but not $I\,
(J^{PC})=0$ or $1\,\; (1^{--})$. When dropping the annihilation
contributions, it is clear in Table III that the state with
quantum numbers $I, J^{PC}=1, 0^{-+}$ has a larger attractive
interaction than that with $I, J^{PC}=1, 1^{--}$, that is consist
with the BES collaboration observation. With the increasing of
annihilation coupling constant, the difference of theses two cases
will emerge.

In summary, in this paper we extend to apply the chiral
constituent quark model with quark-meson-gluon-degrees of the
freedom to the nucleon-antinucleon systems with quantum numbers
$(I, J^{PC})=(1, 1^{--})$; $(1, 0^{-+})$; $(0, 1^{--})$;
$(0,0^{-+})$. Our results with the original model parameters as we
can show the facts as below:

(a) For the $I, J^{PC}=0, 0^{-+}$ ($S$-wave) system, the repulsion
from $\pi$-exchange is so strong that it can cancel the attraction
from $\sigma$-exchange. No matter we adopt the annihilation
coupling constant for the model in a reasonable range, there is no
bound state at all.

(b) The situation for the system with the quantum numbers $I,
J^{PC}=0, 1^{--}$ and $I, J^{PC}=1, 0^{-+}$ for the $S$-wave
nucleon and antinucleon system.

(c) By dynamical calculations we show that the bound states with
$I, J^{PC}=0, 1^{--}; 1, 0^{-+}; 1, 1^{--}$ for an $S$-wave
nucleon and antinucleon system may be exist if the annihilation
coupling constant $\alpha'_s$ is suppressed for certain reason
appropriately. For instance, if letting $\alpha'_s=0.3156$ for
$I,J^{PC}=1, 0^{-+}$, one may obtain a `tightly bound state' at
E=1877.80 MeV and $\sqrt{\langle r^2 \rangle}=2.06 fm$ (see
FIG.6). If we had dropped the annihilation contributions at all,
except $I,J^{PC}=0, 0^{-+}$, in the other three cases ($I,
J^{PC}=1, 1^{--}; 1, 0^{-+}; 0, 1^{--}$) the nucleon and
antinucleon system might be bound tightly within a size not
greater than $1.1$ fm. Moreover, the binding energy of the
$I,J^{PC}=0,1^{--}$ and $I,J^{PC}=1,0^{-+}$ states would be
several tens of MeV greater than that of the state
$I,J^{PC}=1,1^{--}$ and the mean squared root of the radius of the
$I, J^{PC}=1, 1^{--}$ would be a little larger than that of $I,
J^{PC}=1,0^{-+}; 0, 1^{--}$ etc.

In the early 1990s, Dover et al\cite{boundstate} constructed
$\bar{N} N$ potential model from the $N N$ effective potential by
G-parity transformation accordingly, and predicted lower-lying
isospin I=0 natural-parity $J^{PC}=0^{++},1^{--},2^{++}$ bound
states and a few isospin I=1 states, such as $0^{-+},1^{--}$,
also. Their predictions are consistent with our results, if one
compare those of an $S$-wave nucleon and antinucleon system only.
It is interesting to note that there are so many similarities
qualitatively, although Dover's model and ours are based on very
different level, especially, to deal with the short-range behavior
of $\bar{N}N$ in a very different way in the two approaches (in
Dover approach, an arbitrary square-well cutoff is applied for the
unknown short-range behavior of $\bar{N}N$ potential).

Finally, we should note here that this work is very preliminary in
studying the nucleon-antinucleon interaction in the framework of
constituent quark model. There are quite a lot of factors for the
nucleon-antinucleon interaction which should be investigated
carefully, especially, as a very strong assumption, we ignore the
couple channel (such as the multi-pion and other mesons being
involved etc) effects at all, although there are a lot of channels
which should be considered \cite{baryonium}. Furthermore, the
adopted resonating group method also should be tested thoroughly.
Even though, we still would like to emphasize that searching for
the $N\bar{N}$ bound state with the quantum number $I, J^{PC}=1,
0^{-+}$ through various multi-meson decay channels is crucial to
confirm the multi-quark conjecture. Even though considering our
results and the BES observation, we would like to say that, if
there is really an $S$-wave $ p \bar p$ bound state, then its
quantum number is likely to be $I, J^{PC}=1, 0^{-+}$. As a
consequence, similar enhancements in the decays, such as $B^+\to n
\bar{n} K^+,\;B^+\to p\bar{n}K_S;\;$ and $\bar{B^0} \to n\bar{n}
D^0,\; \bar{B^0} \to n\bar{p} D^+\;$ etc in Belle and Babar at
B-factories, and such as $J/\psi \to n\bar{n} \gamma$ etc in BES
at BEPC should be observed, although there are technical
difficulties for the observations.


{\bf Acknowledgements} We thank Zhong-Yie Zhang, Peng-Nian Shen,
Fan Wang and Jia-Lun Ping for useful discussions. This work is
supported by the NSFc under Contract No.10347143, 10247001 and
90103016.


\begin{thebibliography}{99}
\bibitem{bai} J.Z. Bai et al, BES Collaboration, Phys. Rev. Lett. {\bf 91} (2003) 022001.
\bibitem{k1} K. Abe. et al., Phys. Rev. Lett. {\bf 88} (2002) 181803.
\bibitem{k2} K. Abe. et al., Phys. Rev. Lett. {\bf 89} (2002) 151802.
\bibitem{fermi} E. Fermi and C.N. Yang, Phys. Rev. {\bf 76} (1949) 1739.
\bibitem{richard} J.M.Richard, nucl-th/9909030.\\
{\noindent E.Klempt, F.Bradamante, A.Martin, and J.M.Richard, Phys. Rep. {\bf 368} (2002) 119.}
\bibitem{rosner} J.L. Rosner, Phys. Rev. D {\bf 68}(2003) 014004.
\bibitem{zou} B.S. Zou and H.C. Chiang, Phys. Rev. D {\bf 69} (2004) 034004.
\bibitem{datta} A. Datta, P.J. O'Donnell, Phys.Lett.{\bf B567}(2003)273.
\bibitem{xiao} Xiao-gang He and Xue-qian Li, hep-ph/0403191.
\bibitem{paris1}J. Cote, M. Lacombe, B. Loiseau, B. Moussallam, R. Vinh Mau,
Phys. Rev. Lett. {\bf 48} (1982) 1319.
\bibitem{paris2}M. Pignone, M. Lacombe, B. Loiseau, R. Vinh Mau, Phys. Rev. Lett{\bf 67}
(1991) 2423.
\bibitem{paris3}M. Pignone, M. Lacombe, B. Loiseau, R. Vinh Mau, Phys. Rev. {\bf C50} (1994)
2710.
\bibitem{paris4}B. El-Bennich, M. Lacombe, B. Loiseau, R. Vinh Mau, Phys. Rev.
{\bf C59} (1999) 2313.
\bibitem{bonn}V. Mull, K. Holinde, Phys. Rev. {\bf C51} (1995) 2360.
\bibitem{nijmegen1}P.H. Timmers, W.A. van der Sanden, J.J. de swart,
Phys. Rev. {\bf D29} (1984) 1928.
\bibitem{nijmegen2}R. Timmermans, Ph.D. Thesis, The Netherlands, 1991.
\bibitem{nijmegen3}J.J. de Swart, R. Timmermans,
The antibaryon-baryon \\
{\noindent interaction, in: G. Kernel, P. Krizan, M. Mikuz(Eds.),} \\
{\noindent Third Biennial Conference on Low-energy}\\
{\noindent Antiproton Physics: LEAP'94, Bled,Slovenia,12-17}\\
{\noindent September 1994, World Scientific,Singapore, 1995.}
\bibitem{boundstate} W.W. Buck et al. Ann. Phys. {\bf 121} 47
(1979); C.B. Dover et al. Phys. Rev. {\bf D17} 1770 (1978); C.B.
Dover et al. Phys. Rev. {\bf C43} 379 (1991).
\bibitem{zhang}Y.W. Yu, Z.Y. Zhang, P.N. Shen and L.R. Dai, Phys. Rev.
{\bf C52} 1 (1995).
\bibitem{fujiwara}Y. Fujiwara, C. Nakamoto and Y. Suzuki, Phys. Rev. Lett.
{\bf 76} 2242 (1996).
\bibitem{va}F. Fernandez, A. Valcarce, U. Staub and A. Faessler,
J. Phys. G {\bf 19} 2031 (1993).
\bibitem{wang}F. Wang, G.H. Wu, L.J. Teng and T. Goldman, Phys. Rev.
Lett. {\bf 69} 2901 (1992).
\bibitem{zhang2}Q.B. Li, P.N. Shen, Z.Y. Zhang and Y.W. Yu, Nucl. Phys.
{\bf A683} (2001) 487.
\bibitem{georgi}A. Manohar and H. Georgi, Nucl. Phys. {\bf B234} (1984) 189.
\bibitem{pdg} Particle Data Group (PDG), Phys. Rev. D {\bf 66} 010001 (2002).
\bibitem{zhu}Chongshou Gao and Shilin Zhu, hep-ph/0308205.
\bibitem{buchmann}A.J. Buchmann, Y. Yamauchi, A. Faessler,
Nucl. Phys. {\bf A496} (1989) 621
\bibitem{baryonium}L. Montanet, G.C. Rossi,G. Veneziano, Phys. Rep. {\bf 63} (1980) 149.
\end{thebibliography}
\end{document}